\def\be{\begin{equation}}
\def\ee{\end{equation}}
\def\bea{\begin{eqnarray}}
\def\eea{\end{eqnarray}}

\title{Can spacetime curvature be used in future navigation systems?}

\author{Hernando Quevedo\footnote{Email: quevedo@nucleares.unam.mx} \\ 
        Instituto de Ciencias Nucleares, \\ Universidad Nacional Aut\'onoma de M\'exico, \\
					AP 70543, M\'exico, DF 04510, Mexico;\\
Dipartimento di Fisica and ICRANet, \\
Universit\`a di Roma ``La Sapienza", \\ I-00185 Roma, Italy; \\
	Department of Theoretical and Nuclear Physics,  \\ Kazakh National University, \\Almaty 050040, Kazakhstan}

\date{\today}

\documentclass[11pt,english]{article}
\usepackage[]{graphicx}

\begin{document}
\maketitle

\begin{abstract}

We argue that the curvature generated by a gravitational field can be used to calculate the corresponding metric which determines the trajectories of freely falling test particles. To this end, we present a method to compute the metric from a given curvature tensor. We use Petrov's classification to handle the structure and properties of the curvature tensor, and Cartan's structure equations in an orthonormal tetrad to investigate the differential equations that relate the curvature with the metric. The second structure equation is integrated to obtain the explicit expression for the connection $1-$form from which the components of the orthonormal tetrad are obtained by using the first structure equation. This opens the possibility of using the curvature of astrophysical objects like the Earth to determine the position of freely falling satellites that are used in modern navigation systems.

\end{abstract}

\section{Introduction}
\label{sec:int}

One of the most important practical applications of general relativity is the Global Positioning System (GPS), the most advanced navigation system known today.  
It consists essentially in a set of artificial satellites freely falling in the gravitational field of the Earth. To determine the location 
of any point on the Earth by using the method of triangulation, it is necessary to know the exact position of several satellites at a given moment of time. This means that the path of each satellite must be determined as exact as possible. In fact, due to the accuracy expected from the GPS, specially for navigation purposes, it is necessary to take into account relativistic effects for the determination of the satellites trajectories and the gravitational field of the Earth. This method is therefore essentially based upon the use of the geodesic 
equations of motion for each satellite. Moreover, it is necessary to consider the fact that according to special and general relativity clocks inside the satellites run differently than clocks on the Earth surface. Indeed, it is known that not taking relativistic effects into account would lead to an error in the determination of the position which could grow up to 10 kilometers per day.

The curvature seems to be an alternative way to determine the position of any point on the surface of the Earth. Indeed, if we could measure
the curvature of the spacetime around the Earth, and from it the corresponding metric, one could imagine that the determination of the position of the satellites could be carried out in a different way. Maybe this method could be more efficient and more accurate. To this end, it is  necessary to measure the curvature of spacetime. Several devices have been proposed for this purpose. The five-point curvature detector \cite{syn60} consists of four  mirrors and a light source. By measuring the distances between all the components of the detector, it is possible to determine the curvature. Another method uses a local orthonormal frame which is Fermi-Walker propagated along a geodesic 
\cite{pir56}. A gyroscope is directed along each vector of the frame so that the relative acceleration will allow the determination 
of the curvature components. The gravitational compass \cite{sze65}  is a tetrahedral arrangement of springs with test particles on each vertex. Using the geodesic deviation equation, from the strains in the springs it is possible to infer the components of the curvature.
More recently, a generalized geodesic deviation equation was derived which, when applied to a set of test particles, can be used to measure  the components of the curvature  tensor \cite{dirk16}.

It seems therefore to be now well established that the curvature can be measured by using different devices that are within the reach of 
modern technology. The question arises whether it is possible to obtain the metric from a given curvature tensor. This is the problem we will address in this work. In Sec. \ref{sec:mat}, we study a particular matrix representation of the curvature tensor which allows us 
to calculate its eigenvalues in a particularly simple way. Petrov's classification  is used to represent the curvature matrix in terms of its eigenvalues. In Sec. \ref{sec:cartan}, we use Cartan's formalism to derive all the algebraic and differential equations which must be combined and integrated to determine the components
of the metric from the components of the curvature. As particular examples, we present the Schwarzschild, Taub-NUT and Kasner metrics 
with cosmological constant. All the components of the metric are found explicitly in terms of the components of the curvature tensor. It turns out that for a given vacuum solution it is possible to find several generalizations which include the cosmological constant.


\section{Matrix representation of the curvature tensor}
\label{sec:mat}

There are several ways to represent and study the properties of the curvature tensor. Here, we will use a method which is based upon 
the formalism of differential forms and the matrix representation of the curvature tensor. The reason is simple. 
Imagine an observer in a gravitational field. Locally, the observer can introduce a set of four vectors $e_a$ to perform measurements
and experiments. Although it is possible to choose the direction of each vector arbitrarily, the most natural choice would be 
to construct an orthonormal system, i.e., $e_a\otimes e_b = \eta_{ab}= {\rm diag}(+1,-1,-1,-1)$. Of course, the observer could also 
choose a local metric which depends on the point. Nevertheless, the choice of a constant local metric facilitates the process of carrying out
measurements in space and time.  This choice is also in the spirit of the equivalence principle which states that locally it is always 
possible to introduce a system in which the laws of special relativity are valid. The set of vectors $e_a$ can be used to introduce 
a local frame $\vartheta^a$ by using the orthonormality condition $e_a \rfloor \vartheta^b = \delta_a^b$, where $\rfloor$ is the internal product. 
The set of 1-forms $\vartheta^a$ determines a local orthonormal tetrad that is the starting point for the construction of the formalism of differential forms which is widely used in general relativity.

There is an additional advantage in choosing a local orthonormal frame. General relativity is a theory constructed upon the assumption of 
diffeomorphism invariance, i.e, it is invariant with respect to arbitrary changes of coordinates $x^\mu\rightarrow x^{\mu'}$ such that 
$J={\det}\left(\frac{\partial x^{\mu'}}{\partial x^\mu}\right)\neq 0$. Once a local orthonormal frame $\vartheta^a$ is chosen, the only freedom which remains is the transformation 
$\vartheta^a\rightarrow \vartheta^{a'} = \Lambda^{a'}_{\ a} \vartheta^a$, where  $\Lambda^{a'}_{\ a}$ is a Lorentz transformation, satisfying the 
condition $\Lambda^{a'}_{\ a} \Lambda_{a'\, b} = \eta_{ab}$. This means that the diffeomorphism invariance reduces locally to the Lorentz 
invariance, which is easier to be handled. 

In the local orthonormal frame, the line element can be written as
\be
ds^2 = g_{\mu\nu} dx^\mu \otimes dx^\nu= \eta_{ab}\vartheta^a\otimes\vartheta^b\ ,
\ee
with 
\be 
\vartheta^a = e^a_{\ \mu}dx^\mu\ .
\ee
The components $e^a_{\ \mu}$ are called tetrad vectors, and can be used to relate tetrad components with coordinate components. For instance, 
the components of the metric are given in terms of the tetrad vectors by $g_{\mu\nu}=e^a_{\ \mu} e^b_{\ \nu} \eta_{ab}$. 
The exterior derivative of the local tetrad is given in terms of the connection $1-$form $\omega_{ab}$ as \cite{mtw}
\be
d\eta_{ab} = \omega_{ab}+\omega_{ba}\ .
\ee
Since the local metric is constant, the above expression vanishes, indicating that the connection $1-$form is antisymmetric. Furthermore, 
the first structure equation 
\be
d\vartheta^a = - \omega^a_{\ b }\wedge d\vartheta^b\ ,
\ee
can be used to calculate all the components of the connection $1-$form. Finally, the curvature $2-$form is defined as 
\be
\Omega^a_{\ b} = d\omega^a_{\ b} + \omega^a_{ \ c} \wedge \omega^c_{\ b} 
\ee
in terms of a connection. In this differential form representation, the Ricci and Bianchi identities can be expressed as
\be
\Omega^a_{\ b} \wedge \vartheta^b=0\ , \qquad 
d\Omega^a_{\ b} + \omega^a_{\ c}\wedge \Omega^c_{\ b} - \Omega^a_{\ c}\wedge \omega^c_{\ b} =0\ ,
\ee
respectively. 

The curvature $2-$form can be decomposed in terms of the canonical basis $\vartheta^a\wedge \vartheta^b$ as 
\be
\Omega^a_{\ b} 
= \frac{1}{2} R^a_{\ bcd} \vartheta^c\wedge\vartheta^d \ ,
\ee
where $R^a_{\ bcd}$ are the components of the Riemann curvature tensor in the tetrad representation. 

It is well known that the curvature tensor can be decomposed in terms of its irreducible parts which are the Weyl tensor
\cite{deb56}
\be
W_{abcd}=R_{abcd}+2\eta_{[a|[c}R_{d]|b]}+\frac{1}{6} R\eta_{a[d}\eta_{c]b}\ ,
\label{weyl}
\ee
the trace-free Ricci tensor
\be
 E_{abcd} = 2\eta_{[b|[c}R_{d]|a]} - \frac{1}{2}R\eta_{a[d}\eta_{c]b}\ ,
\label{trace}
\ee
and the curvature scalar
\be
S_{abcd} = -\frac{1}{6} R\eta_{a[d}\eta_{c]b}\ ,
\label{scal}
\ee
where we use the following convention for the components of  the Ricci tensor:
\be
R_{ab}=\eta^{cd}R_{cabd}\ .
\ee

Due to the symmetry properties of the components of the curvature tensor, it is possible to represent it 
as a  (6$\times$6)-matrix by introducing the bivector indices $A,B,...$ which encode the information of two different tetrad indices, 
i.e., $ab\rightarrow A$. We follow the convention used  in \cite{mtw} which establishes the following correspondence between tetrad and bivector indices
\be
01\rightarrow 1\ ,\quad 02\rightarrow 2\ ,\quad 03\rightarrow 3\ ,\quad 23\rightarrow 4\ ,\quad 31\rightarrow 5\ ,\quad 12\rightarrow 6\ .
\ee
This correspondence can be applied to all the irreducible components of the Riemann tensor given in Eqs.(\ref{weyl})--(\ref{scal}). Then, 
the bivector representation of the Riemann tensor reads 
\be
R_{AB} = W_{AB} + E_{AB} + S_{AB}\ ,
\label{so31}
\ee
with
\bea
W_{AB}=&\left(
         \begin{array}{cc}
           M & N \\
           N & -M \\
         \end{array}
       \right), \\
       E_{AB}=&\left(
         \begin{array}{cc}
           P & Q \\
           Q & -P \\
         \end{array}
       \right),\\
       S_{AB}=&-\frac{R}{12}\left(
         \begin{array}{cc}
           I_{3} & 0 \\
           0 & -I_{3} \\
         \end{array}
       \right).
\label{so31all}
\eea

Here $M$, $N$ and $P$ are $(3\times 3)$ real symmetric matrices, whereas $Q$ is antisymmetric. 

We see that the bivector representation of the curvature is in fact given in terms of the (3$\times$3)-matrices $M$, $N$, $P$, $Q$ and the scalar $R$, suggesting an equivalent representation in terms of only (3$\times$3)-matrices. 
Indeed, since (\ref{so31}) represents the irreducible pieces of the curvature with respect to the Lorentz group $SO(3,1)$ and, in turn, this group is isomorphic to the group $SO(3,C)$, it is possible to introduce a local complex basis where the curvature is given as a (3$\times$3)-matrix. This is the $SO(3,C)$-representation of the Riemann tensor \cite{deb56,quev92}:
\bea
{\cal R}&=W+E+S\,,\\
W&=M+iN\,,\\
E&=P+iQ\,, \\
S&=\frac{1}{12} R\,I_3\,.
\eea

In this representation, Einstein's equations can be written as algebraic equations. Consider, for instance, a vacuum spacetime for which 
$E=0$ and $S=0$. Then, the vanishing of the Ricci tensor in terms of the components of the Riemann tensor corresponds to the algebraic condition 
\be
{\rm Tr}(W)=0\ , \quad W^T = W\ .
\ee
In general, from  Einstein's equations in the presence of matter
\be
R_{ab} - \frac{1}{2}R\eta_{ab} + \Lambda \eta_{ab} = - \kappa T_{ab} \ ,
\ee
we find that 
\be
R= 4\Lambda + \kappa T\ ,\quad T=\eta^{ab}T_{ab} \ ,
\ee
and the components of the curvature tensor satisfy the relationships
\be
\eta^{cd}R_{cabd} = \kappa T_{ab} + \left( \Lambda + \frac{\kappa}{2} T\right)\eta_{ab}\ .
\ee
It is then easy to see that the following curvature tensor
\be
S=\frac{1}{12}(4\Lambda + \kappa T)\,{\rm diag}(1,1,1)\ ,
\ee
\be
E= \frac{\kappa}{2}\left( 
\begin{array}{ccc}
T_{11} - T_{00} + \frac{1}{2}T & T_{12} - i T_{03} & T_{13} + i T_{02} \\
T_{12} - i T_{03} & T_{22} - T_{00} + \frac{1}{2}T	& T_{23}-i T_{01} \\
T_{13} - i T_{02} & T_{23}+i T_{01} & T_{33} - T_{00} + \frac{1}{2}T
\end{array}
\right)\ ,
\ee

\be
W {\hbox{\ \ arbitrary $(3\times 3)-$matrix with \ \ }} {\rm Tr}(W)=0\ ,\ \ W^T=W\ \ ,
\ee
satisfies  Einstein's equations identically. Thus, we see that the matrix $W$ has only ten independent components, 
the matrix $E$ is hermitian with nine independent components and the scalar piece $S$ has only one component.
 
The energy-momentum tensor determines completely only  the trace-free Ricci tensor and the scalar curvature. 
The Weyl tensor contains in general ten independent components. However, since the local tetrad $\vartheta^a$ 
is defined modulo transformations of the Lorentz group $SO(3,1)$, we can use the six independent parameters
of the Lorentz group to fix six components of the Weyl tensor. Accordingly, we can use the eigenvalues 
of the matrix $W$ to write the four remaining parameters in the form
\be
W^{^I}=\left(
\begin{array}{ccc}
a_1+ib_1 & & \\
 & a_2+i b_2 & \\
 & & -a_1-a_2 -i(b_1+b_2) 	
\end{array}
\right)\ .
\ee
In fact, this is the most general case of a Weyl tensor, and corresponds to a type $I$ curvature tensor in Petrov's classification.
If the eigenvalues of the matrix $W$ are degenerate, then $a_2=a_1=a$ and $b_2=b_1=b$ and therefore
\be
W^{^D} = \left(
\begin{array}{ccc}
a+ib & & \\
 & a+i b & \\
 & & -2 a -2 i b  	
\end{array}
\right)\ ,
\ee
which represents a type $D$ curvature tensor. 

In general, all the eigenvalues can depend on the coordinates $x^\mu$ of the spacetime. The real part of the eigenvalues $a_1$ and $a_2$ 
represent the gravitoelectric part of the curvature, whereas the imaginary part $b_1$ and $b_2$ correspond to the gravitomagnetic field, i.e.,
the gravitational field generated by the motion of the source.


\section{Integration of Cartan's structure equations}
\label{sec:cartan}

Our aim now is to show that for a given curvature tensor it is possible to integrate Cartan's equation in order
to compute the components of the metric. To this end, it is necessary to rewrite Cartan's equations so that the dependence on 
the spacetime coordinates becomes explicit. First, let us introduce the components of the anholonomic connection $\Gamma^a_{\ bc}$ 
by means of the relationship
\be
\omega^a_{\ b} = \Gamma^a _{\ bc} \vartheta^c\ ,
\ee 
and the condition $\Gamma_{abc}=-\Gamma_{bac}$. Then, from the definition of the connection $1-$form, we obtain
\be
e_{\ [\mu,\nu]}^{a} = \Gamma^a_{\ bc} e_{[\nu}^{\ \ b} e_{\mu]}^{\ \ c} \ ,
\label{eq1}
\ee
which represents a differential equation for the components of the tetrad vectors $e^a_{\ \mu}$. Here, the square brackets denote antisymmetrization. On the other hand, the exterior derivative of the curvature $2-$form yields
\be
d\Omega^a_{\ b} =\frac{1}{2} \left(R^a_{\ bcd,\mu } e^{\ \mu}_{e} + 2 R^a_{\ bfd} \Gamma ^f_{\ ec}\right) \vartheta^e\wedge \vartheta^c
\wedge\vartheta^d \ ,
\ee
which together with
\be
d\Omega^a_{\ b} = \frac{1}{2}\left(R^f_{\ bed}\Gamma^a_{\ fc}\right) \vartheta^e\wedge\vartheta^c\wedge\vartheta^d \ ,
\ee
leads to the following equation
\be
R^a_{\ b[cd,|\mu|}e^{\ \mu}_e = R^a_{\ f[cd}\Gamma ^f_{\ |b|e]} - R^a_{\ b[cd}\Gamma^f_{\ |f|e]} - 2 R^a_{\ bf[c}\Gamma^f_{\ de]}\ .
\label{eq2}
\ee
This equation represents an algebraic relationship between the components of the tetrad vectors $e^a_{\ \mu}$ and the components of 
the connection $1-$form $\Gamma^a_{\ bc}$. 

Finally, the components of the curvature tensor can be expressed in terms of the anholonomic components of the connection as
\be
\frac{1}{2}R^a_{\ bcd} = \Gamma^a_{\ b[d,|\mu|}e_c^{\ \mu} + \Gamma^a_{\ be}\Gamma^e_{\ [cd]} + \Gamma^a_{\ e[c}\Gamma^e_{\ |b|d]}\ ,
\label{eq3}
\ee
which can be considered as a system of partial differential equations for the components of the connection with the components of the curvature and the tetrad vectors as variable coefficients.

To integrate Cartan's equations we proceed as follows. First, we consider the 20 particular independent equations (\ref{eq3}) together with the 18 equations which follow from Eq.(\ref{eq2}). The idea is to obtain from here all the 24 anholonomic components of the connection
 $\Gamma^a_{\ bc}$. Then, this result is used as input to solve the 24 independent equations which follow from Eq.(\ref{eq1}). This procedure
leads to a large number of equations which are complicated to be handled. They have been analyzed with some detail in \cite{quev92}.
Here, we will limit ourselves to quoting the some of the final results obtained previously. 

\section{Type $D$ metrics}
\label{sec:metd}

Consider a type $D$ curvature tensor with eigenvalue $a+ib$, and suppose that 
\be
a=a(x^3)\ ,\quad b=b(x^3) \ ,
\ee
i.e., we assume that the curvature depends on only one spatial coordinate. Furthermore, it is well known that type $D$ spacetimes can have
a maximum of four Killing vector fields. Then, we will consider spacetimes with two Killing vector fields which can be taken along the coordinates $x^0$ and $x^1$; consequently,
\be
g_{\mu\nu,0} = g_{\mu\nu,1}=0\ , \quad g_{\mu\nu,0} = \frac{\partial g_{\mu\nu}}{\partial x^0}\ .
\ee
This means that the only relevant spatial direction should be $x^3$. Therefore, we can use the diffeomorphism invariance of general relativity in order to bring four metric components into any desired form. We then assume that
\be
g_{30}=g_{31}=g_{32}=0\ ,\quad g_{33} = g_{33}(x^3)\ .
\ee

In terms of the local tetrad, the above assumption implies that
\be
\vartheta^3 = \sqrt{g_{33}} dx^3 =  e^3_{\ \dot 3} dx^3 \ ,
\ee
where the dot denotes coordinate indices.  As a consequence we have that
\be
d\vartheta^3 = 0 \ ,
\ee
which implies that six components of the tetrad vectors vanish, namely,
\be
e^0_{ \ \dot 3} = e^1_{ \ \dot 3} =  e^2_{ \ \dot 3} = e^3_{\ \dot 0} = e^3_{\ \dot 1}=e^3_{\ \dot 2}\ .
\ee
This means that we now have a system of only ten components of $e^a_{\ \mu}$ that are unknown. On the other hand,
the vanishing of the exterior derivative of $\vartheta^3$ implies that
\be
\Gamma^3_{\ [ab]} = 0 \ ,
\ee
which drastically simplifies the set of differential equations for the components of the connection. 
A detailed analysis of the resulting equations shows that it is convenient to consider particular cases 
which are obtained for different choices of some components of the connection. In fact, it turns out that 
the choices 
\be
\Gamma^1_{\ 21} = 0 \ , \quad \Gamma^1_{\ 23}\neq 0 
\ee
and
\be
\Gamma^1_{\ 21} \neq 0 \ , \quad \Gamma^1_{\ 23} =  0 
\ee
lead to completely different solutions which we will analyze in the following subsections.

It is then possible to 
show that with these simplifying assumptions, we can integrate the set of partial differential equations. Several 
arbitrary functions arise in the tetrad vectors which can then be absorbed by means of coordinate transformations.

\subsection{Schwarzschild and Taub-NUT metrics}

The particular choice 
\be
\Gamma^1_{\ 21} \neq 0 \ , \quad \Gamma^1_{\ 23} =  0 
\ee
leads to a compatible set of algebraic and differential equations which allow us to calculate all the components of the tetrad vectors.
We present the final results without the details of calculations which can be consulted in \cite{quev92}.

Consider, for instance, the following curvature tensor in the $SO(3,C)$ representation:
\be
{\cal R} = -\frac{M}{r^3} \, {\rm diag}(1,1,-2)+ \frac{\Lambda}{3}\,{\rm diag}(1,1,1)\ ,
\ee
where $r=x^3$. Then, the integration of all the differential equations yields
\be
e^3_{\ \dot 3}\left(\alpha -\frac{2M}{r} - \frac{\Lambda}{3} r^2\right)^{-1/2}\ ,\quad e^2_{\ \dot 2} = r\ ,\quad 
e^1_{\ \dot 2}= r F^1_{\ \dot 2} \ ,
\ee
\be
e^0_{\ \dot m} = C^0_{\ \dot m} \left(\alpha -\frac{2M}{r} - \frac{\Lambda}{3} r^2\right)^{1/2}\ ,\quad
e^0_{\ \dot 2} = F^0_{\ \dot 2} \left(\alpha -\frac{2M}{r} - \frac{\Lambda}{3} r^2\right)^{1/2}\ ,
\ee
where $m=0,1$, $\alpha$ and $C^0_{\ \dot m}$ are arbitrary real constants and $F^0_{\ \dot 2}$ and $F^1_{\ \dot 2}$ are non-zero functions
of the coordinate $x^2$. It is then possible to find a coordinate system in which the above tetrad vector components 
lead to the line element
\be
ds^2= \left(\alpha -\frac{2M}{r} - \frac{\Lambda}{3} r^2\right)dt^2 
-\frac{dr^2}{\alpha -\frac{2M}{r} - \frac{\Lambda}{3} r^2} - r^2(d\theta^2 +\sin^2\theta d\phi^2) \ ,
\ee
which represents the Schwarzschild-de-Sitter spacetime.
 
Consider now a curvature tensor with gravitoelectric and gravitomagnetic components:
\be
{\cal R} = - \frac{M+iP}{(r+iC)^3} \, {\rm diag}(1,1,-2) + \frac{\Lambda}{3} \, {\rm diag}(1,1,1)\ ,
\ee
where $P$ and $C$ are arbitrary real constants. It is then possible to show that the result of the integration leads to a 
line element of the form 
\be
ds^2 = \Delta_1 (dt + 2 C \cos\theta d\phi)^2 - \frac{dr^2}{\Delta_1} 
- (r^2+C^2)\left(\Delta_2 \sin^2\theta d\phi^2 + \frac{d\theta^2}{\Delta_2}\right)  \ ,
\ee
with 
\be 
\Delta_1 = (r^2+C^2)\left[\frac{P}{C} (r^2-C^2) - 2 M r - \frac{\Lambda}{3}(r^2+C^2)^2\right]^{-1}\ ,
\ee
\be
\Delta_2 = \frac{P}{C} + \frac{4}{3} \Lambda C^2 \ .
\ee

Different choices of the parameters $P$ and $C$ lead to different particular solutions of Einstein's equations. For instance, the choice
\be
P=l\left(1-\frac{4}{3} \Lambda l^2\right)\ ,\quad C=l
\ee
corresponds to the Taub-NUT metric with cosmological constant \cite{dem72}, where $l$ is the NUT parameter. 
Furthermore, the choice 
\be
P=k l \left(1 -\frac{4}{3}\Lambda l^2\right)\ , \quad C=l\ , \quad k=-1,0,+1
\ee
is known as the Cahen-Defrise spacetime \cite{cd68}. 

The Taub-NUT metric is obtained for the choice $P=l$ and $C=l$ with $\Lambda=0$. It is then possible to obtain several 
different generalizations which include the cosmological constant. In fact, the simplest choice corresponds to $P=C=l$
and the cosmological constant entering only the scalar part of the curvature. Other generalizations are obtained by 
choosing the free parameter $P$ as a polynomial in $\Lambda$, for instance,
\be
C=l\ ,\quad P = l\left(c_1 + c_2 \Lambda l^2 + c_3 \Lambda^2 l^4 + ...\right)\ ,
\ee
where $c_1$, $c_2$, etc. are dimensionless constants. Another example is obtained for the choice
\be
P=l\ ,\quad C = l\left(c_1 + c_2 \Lambda l^2 + c_3 \Lambda^2 l^4 + ...\right)\ .
\ee
All these examples generalize the  Taub-NUT metric to include the cosmological constant. In principle, all of them should represent
different physical configurations since they all differ in the behavior of the Weyl tensor. This opens the possibility of analyzing
anti-de-Sitter spacetimes which are equivalent from the point of view of the scalar curvature, but different from the point 
of view of the Weyl curvature. 

We conclude that in the particular case analyzed here the method presented above can be used to generate
new solutions of Einstein's equations with cosmological constant.

\subsection{Generalized Kasner metrics}
\label{sec:kas}

Another particular choice of the connection components given by 
\be
\Gamma^1_{\ 21} = 0 \ , \quad \Gamma^1_{\ 23} \neq   0 
\ee
leads to a set of algebraic and differential equations which can be integrated completely for a curvature tensor with only 
gravitomagnetic components, i.e.,
\be
{\cal R} = a(x^3) \, {\rm diag}(1,1,-2) + \frac{\Lambda}{3}\, {\rm diag}(1,1,1)\ .
\ee
Indeed, after applying a series of coordinate transformations, the corresponding line element can be expressed as
\bea
ds^2 = & \frac{|a_0|}{\left|3a_0-\frac{\Lambda}{2}\right|^{2/3}} dt^2 
- \frac{(a_0')^2}{2|a_0|\left(3 a_0-\frac{\Lambda}{2}\right)^2}dr^2 \nonumber \\
& - \frac{1}{\left|3 a_0-\frac{\Lambda}{2}\right|^{2/3}}(dX^2 + dY^2) \ ,
\eea
with 
\be 
a_0 = a+ \frac{\Lambda}{2} < 0\ ,
\ee
and the prime represents derivation with respect to $r=x^3$. Here we see that the metric can be calculated immediately from 
the gravitoelectric component of the curvature $a(r)$. Several particular metrics can be written down. We quote only the metric that
follows from the eigenvalue
\be
a = -\frac{\gamma}{r^{2\beta}} - \frac{\Lambda}{3}\ ,
\ee 
where $\gamma$ and $\beta$ are real constants. For this case, we obtain
\bea 
ds^2 = & \left(\gamma  - \frac{\Lambda}{6} r^{2\beta}\right) r^{-2\beta/3} dt^2 
- \frac{2}{9} \beta^2 \left(\gamma  - \frac{\Lambda}{6} r^{2\beta}\right)^{-1} r^{2(\beta-1)} dr^2 \nonumber \\
& - r^{4\beta/3} (dX^2+dY^2) \ .
\label{kas1}
\eea
In the limiting case $\Lambda=0$, we obtain for each value of $\beta$ a particular case of the Kasner metric \cite{solutions}.
In general, the above line element represents a generalization of the Kasner space which includes the cosmological constant.
We see that in this particular case we have chosen a curvature eigenvalue which contains the cosmological constant explicitly.
This has been done in order to obtain a simple expression for the Kasner metric with $\Lambda$. However, one can always change 
in the function $a(r)$ the term containing the cosmological constant, in order to obtain different solutions. The simplest spacetime would correspond
to the one in which the Weyl tensor does not depend on the cosmological constant, i.e.,
\be
a= -\frac{\gamma}{r^{2\beta}}\ ,
\ee
for which we obtain the generalized Kasner metric
\bea
ds^2 = & \frac{\left|-\frac{\gamma}{r^{2\beta}} + \frac{\Lambda}{2}\right|}{\left|-\frac{3\gamma}{r^{2\beta}} + \Lambda\right|^{2/3} } dt^2
- \frac{2\beta^2\gamma^2}{r^{2(2\beta+1)} \left| -\frac{\gamma}{r^{2\beta}} + \frac{\Lambda}{2}\right| 
\left(-\frac{3\gamma}{r^{2\beta}} + \Lambda\right)^2} dr^2 \nonumber \\
& - \frac{1}{\left|-\frac{3\gamma}{r^{2\beta}} + \Lambda\right|^{2/3} }\left(dX^2+dY^2\right)  \ .
\eea
This particular choice seems to be more complicated than the solution (\ref{kas1}); however, from a physical point of view it corresponds to the simplest choice in which the Weyl tensor is not affected by the presence of the cosmological constant.

We see that it is possible to obtain several generalizations of the Kasner metric with cosmological constant and, in principle, each of them
should correspond to a different physical configuration.
 
\section{Conclusions}
\label{sec:con}

In this work, we presented a method to calculate the components of the metric tensor from the components of the Riemann curvature tensor. 
We use the formalism of differential forms and Cartan's structure equations in order to calculate explicitly the algebraic and differential 
equations that relate the components of the local tetrad vectors with the components of the connection $1-$form and the curvature $2-$form.

We integrate the differential equations for the case of a type $D$ curvature tensor in Petrov's classification which is characterized by only one complex eigenvalue. We found that for a given 
curvature eigenvalue, it is possible to obtain different metrics, depending on some assumptions made for the components of the connection 
$1-$form. For the computation of explicit examples, we assume that the curvature eigenvalue depends on only one spatial coordinate.
This simplifies the set of differential  equations and allows us to carry out the integration completely. We obtain as concrete 
examples two classes of spacetimes. The first class contains the Schwarzschild metric, the Taub-NUT metric and several generalizations which include the cosmological constant. The second class contains a family of particular Kasner spacetimes with cosmological constant.

The main result of the present work is that is possible to obtain the metric from the curvature. Furthermore, we found that for any given 
vacuum spacetime, we can apply the procedure presented in this work to obtain different generalizations which include the cosmological constant. This means that solutions of Einstein's equations with cosmological constant are not unique. The main physical difference between 
different spacetimes with cosmological constant is reflected in the Weyl tensor which behaves differently for each metric.

The concrete examples of curvature analyzed in this work involve terms with gravitoelectric monopole (mass parameter) and gravitomagnetic mono\-pole (NUT parameter) only. In the case of an astrophysical gravitational source, a more realistic situation involves higher mass and angular momentum multipole moments. It is then easy to 
see that if we consider the Weyl tensor in the form 
\be
W= - \sum_{n=1}^\infty \frac{m_n}{r^{2n+1} }   \, {\rm diag}(1,1,-2) \ ,
\ee
the integration of the structure equations can be performed in a way similar to the one used to obtain the Schwarzschild and the Taub-NUT 
metrics. The explicit metric components can be computed by using the general formula presented here and in \cite{quev92}. 
The resulting metric will contain the parameters $m_n$ which correspond to higher mass multipole moments. 
In this way, one could
generate exact solutions with a prescribed set of multipoles. In a realistic situation, for instance in the case of the Earth, one would need only a limited number of moments $m_n$, whose values can be from the measurement of the curvature components.

To take into account higher gravitomagnetic moments, it will be probably necessary to generalize the method presented here. Indeed, the presence of rotational moments implies that the curvature must depend on at least two spatial coordinates (a radial and an angular 
coordinate). In addition, it will probably necessary to consider not only type $D$, but also type $I$ Weyl tensors. In this case, we 
need to construct a more general method than the one presented here. However, if we fix the angular coordinate and consider, for instance,
the equatorial plane of the gravitational source, the curvature will depend only on the radial distance and it will be possible to consider
a Weyl tensor of the form
\be
W= - \sum_{n=1}^\infty \frac{m_n+i j_n}{r^{2n+1} }   \, {\rm diag}(1,1,-2) \ ,
\ee 
where the parameters $j_n$ represent the multipoles of the curvature generated  by the rotation of the source. Of course, this would be 
only an approximation of a realistic compact object since the dependence on the angular coordinate is completely neglected. However, since
in the case of an object like the Earth, the deviations from spherical symmetry due to rotation are very small,  one could expect that
this equatorial plane approximation would lead results with a good degree of accuracy.

We conclude that the method presented in this work can be used, in principle, to generate particular metrics, describing the gravitational field of realistic compact objects.  It would be interesting to investigate this problem in detail in the case of the Earth, 
to study the possibility of developing new navigation systems by using as input the curvature of the spacetime around our planet.

\section*{Acknowledgements}

This work has been supported by the UNAM-DGAPA-PAPIIT, Grant No. IN111617.

\bibliographystyle{unsrt}

\bibliography{quevedo_relgeo_proceedings_2016}

\begin{thebibliography}{10}

\bibitem{syn60}
J.~L. Synge.
\newblock {\em Relativity: The general theory}.
\newblock North-Holland, Amsterdam, 1960.

\bibitem{pir56}
F.~A.~E. Pirani.
\newblock {\em Acta Phys. Pol.}, 15:389, 1956.

\bibitem{sze65}
P.~Szekeres.
\newblock {\em J. Math. Phys.}, 6:1387, 1965.

\bibitem{dirk16}
D.~Puetzfeld and Y.~N. Obukhov.
\newblock {\em Phys. Rev. D}, 93:044073, 2016.

\bibitem{mtw}
K.~S.~Thorne C.~W.~Misner and J.~A. Wheeler.
\newblock {\em Gravitation}.
\newblock W. H. Freeman, San Francisco, USA, 1973.

\bibitem{deb56}
R.~Debever J.~Geheniau.
\newblock {\em Bull. Acad. Roy. Soc. Belgique}, 42:114, 1956.

\bibitem{quev92}
H.~Quevedo.
\newblock {\em Gen. Rel. Grav.}, 24:693, 1992.

\bibitem{dem72}
M.~Demia\'nski.
\newblock {\em Phys. Lett. A}, 42:157, 1972.

\bibitem{cd68}
M.~Cahen and L.~Defrise.
\newblock {\em Commun. Math. Phys.}, 11:16, 1968.

\bibitem{solutions}
M.~MacCallum C.~Hoenselaers H.~Stephani, D.~Kramer and E.~Herlt.
\newblock {\em { Exact solutions of Einstein's field equations}}.
\newblock Cambridge University Press, Cambridge, UK, 2003.

\end{thebibliography}

\end{document}